\documentstyle{mn}

\newcommand{\mincir}{\ \raise -2.truept\hbox{\rlap{\hbox{$\sim$}}\raise5.truept
	\hbox{$<$}\ }}			
\newcommand{\magcir}{\ \raise -2.truept\hbox{\rlap{\hbox{$\sim$}}\raise5.truept
 	\hbox{$>$}\ }}

\begin{document}

\title[A Primeval galaxy at z \magcir 4.4 ]{The optical identification
of a primeval galaxy at {\it z} $ \magcir 4.4$ 
\thanks{Based on observations collected at the European Southern Observatory,
La Silla, Chile}
}

\author[A. Fontana et al.]
{
Adriano Fontana$^1$, Stefano Cristiani$^{2,3}$, Sandro
D'Odorico$^3$, Emanuele Giallongo$^1$ \cr
and Sandra Sa\-va\-glio$^{3,4}$\\
$^{1}$ Osservatorio Astronomico di Roma, 00040 Monte Porzio, Italy \\
$^{2}$ Dipartimento di Astronomia dell' Universit\`a, 35122
Padova, Italy\\
$^{3}$ 
European Southern Observatory, 85748 Garching bei M\"unchen, Germany\\
$^{4}$ Istituto di Astrofisica Spaziale del CNR, 00044 Frascati, Italy\\
}
\maketitle
\begin{abstract}

We have  obtained with the SUSI CCD camera on the ESO 3.5m NTT
  deep images  in the BVRI bands of the field centered on
the QSO BRI 1202-0725 ($z_{\it em}=4.694$).
In the final combined frames the stellar images have FWHM of 1,1,0.6 and
0.65 arcsec respectively. The R and I images 
 show clearly a galaxy $2.2''$ from the QSO, corresponding
 to $13h^{-1}_{50}$ kpc at $z\sim 4.5$. Possible identification
with the metal absorption systems seen in the line of sight to the QSO,
including the
highest
 redshift damped
system
known to date at $z=4.383$,  
are discussed. We conclude that 
its colours  can be reconciled only 
with the spectrum of a primeval galaxy
at  $z \magcir 4.4 $, making it the most distant galaxy
detected so far. From its magnitudes and models of young galaxy
evolution
we deduce that
it is forming stars at a rate $\sim 30 M_{\odot}$ yr$^{-1}$ 
and has an estimated age of the order of $10^8$ yr or less, implying
that the bulk of the stellar population formed at $z < 6$.
\end{abstract}
\nokeywords
\section{Introduction}
The detection of primeval galaxies (PG's), i.e.
high-redshift, young galaxies undergoing their first episodes of star 
formation, is essential to provide the observational
underpinning to modern theories of galaxy formation and evolution.
A strong hint on their existence  
up to redshifts $\approx4$ is provided
by the observation of metal enriched absorption systems in the spectra
of high-redshift QSOs,  but their actual detection  at 
$z \geq 1$ has proved  to be a difficult task.
Searching for the optical counterparts of powerful radio-sources 
has led to the discovery of a number of high redshift galaxies \cite{MCa,LM},
but these objects are hardly informative on the general population of
PGs
because the continuum is contaminated by non-stellar processes and the
emission fluxes and relative intensities are not those of standard
star
formation regions.
Deep spectroscopic surveys at optical wavelengths (e.g. Glazebrook  et al. 
1995a,b, Crampton et al. 1995),
have led to the identification of galaxies typically at redshift 
up to $z\sim 1$.
Field surveys based on the search of highly redshifted
Lyman-$\alpha$ 
line emission or other emission lines as a signature of star formation
have been equally unsuccessful at large redshifts (Djorgovski et al. 1993,
 Pahre \& Djorgovski 1995).

An alternative searching technique is based on ultra-deep imaging in
different colours in the fields of high redshift QSOs which show
Lyman Limit and Damped Lyman-$\alpha$ Absorption Systems (DLAS) 
in their absorption spectra.
This method has been successful in a few cases (Steidel \& Hamilton 1992, 
Steidel et al. 1995, Giavalisco et al. 1994), detecting  galaxies
at redshift $z=3- 3.5$.

We have applied this approach to study the field of
the 18.7 R magnitude,$z=4.694$ QSO BRI 1202-0725. The quasar was found
during the APM optical survey for QSO with $z>4$ \cite {IMH}
and its coordinates are given in McMahon et al(1994).
Giallongo et al. (1994) obtained a
spectrum of this QSO  at $\approx 40$ km s$^{-1}$  
and detected a DLAS at $z\approx 4.38$, the highest redshift system of
this
type known to date. Lu et al. (1995) and Wampler et al .(1995) have studied
the metal absorption lines associated to this DLAS and both derived metal
abundances 
(for Fe and O respectively) which are of the order of
$10^{-2}$ of the solar value. 

\section{OBSERVATIONS AND ANALYSIS}
The field of BRI 1202-07 was  observed during three
photometric nights, 23--26 Apr 1995, with the SUSI direct CCD camera
at the Nasmyth focus of the ESO NTT.
Four broad--band filters were used corresponding to
the $B V I$ passbands of the Johnson-Kron-Cousins system (JKC) and to
the $r$ of the Thuan and Gunn system. In addition a narrow band (67 {\AA)
filter centered at $\lambda =6560$ {\AA}, which includes the Lyman-$\alpha$
emission at the $z=4.38$ DLA system, has been used. Several dithered
images were obtained in the four broad bands and in the narrow filter.
The total integration times are 15000,10800,7200,7800 and 7800 seconds
for
the four broad bands and the narrow filter observations respectively.
 The FWHM of
the
stellar images in the combined frames are 1,1,0.6, 0.65 and 1 arcsec. 

The photometric calibration was obtained from several standard stars
observed on the same nights and at similar airmasses. 
Total magnitudes have been obtained 
in an aperture of fixed size ($1.8''$), applying
a seeing--dependent aperture correction.
 
In view of the spectral peculiarities expected for PG's, 
we have calibrated $B' V' R' I'$ magnitudes in the ``natural''
system defined by our instrumental passbands.

The zero-points of our instrumental system have been
adjusted to give the same $BVRI$
magnitudes of the JKC system for stellar objects with
$B-V = V-R = R-I = 0$. 
The colour transformations for stellar objects turn out to be 
$I'=I$, $B'-V' = 0.91\times (B-V)$, $V'-R' = 1.17 \times (V-R)$, 
$R'-I' = 0.74 \times (R-I)$.
The magnitudes have been corrected for galactic absorption
which at the position of the QSO is E(B-V)=0.03.

Figure 1 shows the region surrounding BRI 1202-0725 in the $B'V'R'$ bands. 
A faint, non symmetric galaxy clearly stands out in the $R'$ and $I'$ frames
NW of the QSO at a $P.A.=328^{\circ}$ and a distance of $2.2''$,
corresponding \footnote{$\Omega =1$ and H$_o$
=50 km s$^{-1}$ Mpc$^{-1}$ are adopted throughout the paper}
 to 18--13 kpc  in the redshift range $z=1-4.7$.
Its magnitudes, measured after the QSO subtraction, 
are: $B' \geq 27.5$, $V'=26.2\pm 0.4$, $R'=24.3\pm 0.1$, 
$I'=24.1\pm 0.2$. The lower limit on $B'$ is a $1\sigma$ limit. 
From the narrow band frame (7800s) we set a $1\sigma$ flux upper limit
$f_{NB}< 6.7\times 10^{-17}$ erg s$^{-1}$ cm$^{-2}$.

\begin{figure*}
\vspace{22cm}   
\caption{
Upper left: a $1^{h}$ Gunn-$r$ combined image of the field 
of the QSO
BRI1202-0725. This combination of the single frames of best image quality
 has an image quality of $0.5''$ FWHM.
Lower left and lower right: combined images in the $B'$ ($15000^s$), and $V'$ ($10800^s$)
bands, respectively, both with $1''$ FWHM . 
Upper right: a $2^{h}$ Gunn-$r$ combined image (FWHM $0.65''$) in which a scaled
stellar PSF has been subtracted to the QSO and to the star in the
upper right corner. The galaxy angular size is 1.7''.
The field is $22''\times 22''$, North is at the
top, East to the left.
}
\label{fig1}
\end{figure*}

This galaxy has also been independently detected in deep K band exposures at
the Keck telescope (Djorgovski 1995), with an  estimated 
K magnitude of $23 \pm 0.5$ (Djorgovsky, private communication).

\section{REDSHIFT IDENTIFICATION}
In discussing the possible redshift of the galaxy we have detected,
we must take into account that the line of sight to the QSO is likely 
to be populated by many groups
of
galaxies at
different redshifts. This can be inferred by the presence of several
metal absorption systems. Beside the DLAS at $z=4.383$ shown by
Giallongo et al (1994),
Wampler et al. (1995)
discuss two systems at
$z=4.672$
and $z=4.687$  and list 6 other metal systems with redshift between
1.75 and 4.48. 

The  peculiar colours of the detected galaxy, $B'-R' > 3.1$, $V'-R'=1.9$,
$R'-I'=0.2$, $R'-K=1.3$, agree with those expected for 
a $z\approx 4.4$ star forming 
galaxy, whose UV radiation from
young massive stars is redshifted into the optical bands.
The characteristic spectral features of such an object are, in fact,
the Lyman continuum  break at 912 {\AA} (rest--frame)
and a relatively flat spectrum longward of it (Bruzual \& Charlot 1993). 
Additional attenuation
of the emitted flux shortward of 1215 {\AA} (rest--frame) occurs because 
of the absorption by the population of intergalactic Lyman-$\alpha$
clouds.
An interesting comparison can be made with a spectrum of a PG
with constant star formation rate (SFR) and an age of 0.1 Gyr kindly
computed and redshifted at 4.4 by Madau according to his standard model
(Madau 1995).
Convolving this synthetic spectrum (Fig. 2a) with our
instrumental passbands and normalizing the galaxy flux to $I'=24.1$,
we derive the following magnitudes:
$R'=24.5$, $V'=25.9$, and $B'=29.8$, consistent with the observations. 
This agreement implies a low Lyman-$\alpha$ emission, which
 is consistent with the
upper limit on the Lyman-$\alpha$  equivalent  width
$ <150$ {\AA} derived from the narrow band frame.
\begin{figure*}
\begin{minipage}{150mm}
\vspace{100mm}
\caption{
{\it a)}
Comparison between a model spectrum of a 
primeval galaxies at $z=4.38$
and the observed fluxes. The dotted line shows the intrinsic spectrum,
the solid line
the spectrum with the intergalactic
Lyman absorption included. 
The upper horizontal axis shows the rest frame wavelengths. 
The observed flux is shown on the left vertical axis and
 the rest--frame luminosity on the right one. 
The conversion from our optical
magnitudes into monochromatic fluxes has been calibrated on standard stars with
no correction  for the peculiar spectrum of this galaxy.
The model spectrum has been normalized to the observed $I$ flux. 
Fluxes have been plotted at the peak of filter transmission, and 
the errorbars in the x direction show the FWHM of the filters.
{\it b)} Same as 
{\it a}, 
but for a primeval galaxy at $z=4$
(solid line) and one at $z=4.7$ (dotted line), both with the intergalactic
Lyman absorption included. 
}
\label{fig2}
\end{minipage}
\end{figure*}

At the same time, 
the colours  are not compatible with those of a galaxy at significantly lower
redshift.
In the case of a galaxy that is actively forming stars,
the UV radiation emitted at wavelengths  longward of its Lyman break 
moves in the bluer bands  as the redshift decreases, resulting
in $V'-R'$ and $B'-R'$ colours much flatter than those observed. 
For instance a 0.7 Gyr old, star--forming galaxy  
would invariably have (De Robertis and McCall 1995) $V-R < 0.5$, and
$B-V < 0.5$ in the redshift range $1<z<3$.
In the case of 
an old stellar population at intermediate redshifts,
of which the large  $V'-R'$ colour might be an indication, 
population synthesis models (De Robertis \& McCall 1995, Bressan et al. 1994)
 predict a much larger $R-I$ colour than 
observed in our PG candidate. According to the models, in fact,
in the redshift interval 1--3 the $V-R$ colour varies between 2 and 2.5, 
but the $R-I$ colour is larger than $ 1.5$.
These arguments apply to the metal systems detected at  $z<3$, 
 which then cannot be associated with the detected galaxy.
In particular, at z=1.7 an active (passive) galaxy would 
result (De Robertis \& McCall 1995, Bressan et al. 1994)
in  $V-R=0$ ($2.1$) and $R-I=0.3$ ($2.1$).

We finally note that the faint $B'$ and $V'$ magnitudes 
of this galaxy cannot be interpreted as due to a strong reddening in a low
redshift object. The  observed
$R'-I'=0.2$ and 
$I'-K\approx 1$ colours would imply, in presence of a substantial
amount of dust, an intrinsic spectrum with abnormally negative $R'-I'$ and
$I'-K$ colours.

The observed colours can also be used to discriminate between 
galaxies at 
redshifts around $4.5$.
At these redshifts, even a small difference in redshifts produces 
significant variations of the observed colours (see fig. 2b), due to 
changes of the Lyman forest absorption in the various passbands.
At redshifts lower than 4.4 the $V'$ flux would considerably increase, while at
higher redshifts, the flux in the $R'$ band would
be attenuated by the Lyman forest.
The same PG that at $z\sim4.4$ is consistent with the
observations, at $z=4$ (with the same flux normalization in the I band) would 
still have $R'=24.3$, but $V'=25.3$, i.e. 0.9 mag. brighter than observed.
At $z=4.7$ (with the same intergalactic absorption as in BRI
1202-0725), it 
would  have $V'=26.5$ but $R'=25.1$, i.e. 0.8 mag. 
fainter than observed in the $R$ band.
Nevertheless, a Lyman-$\alpha$ emission
might increase the $R'$ flux and reconcile the predictions
with the observations.
At $z=4.7$, the required strength of
such a line, causing a $\Delta R'=-0.8$ mag.,  is
$f_{Ly\alpha}\sim2\times 10^{-16}$ erg s$^{-1}$ cm$^{-2}$,
still lower by an order of magnitude 
than the upper limit inferred from the negative detection of the [O II]
3727 {\AA} line in the K band (Pahre \& Djorgovsky 1995).
Redshifts much 
higher than 4.7 can be excluded, since the  Lyman-$\alpha$ emission
would shift out of the $R'$ filter.

We conclude that we have detected a 
galaxy that is massively forming stars at $z = 4.4 - 4.7$:
our deduction
is based on what it is presently known or reasonably assumed on galaxy
evolution and only a spectroscopic confirmation will provide the
ultimate evidence that our argument is correct.

\section{DISCUSSION}
The photometric properties of this galaxy can be 
compared with  standard spectral evolution
 models (Bruzual \& Charlot 1993, Charlot \& Fall 1993)
to infer its evolutionary status:
the  UV luminosity is related to the instantaneous SFR occurring
in the galaxy, while the overall shape of the spectrum  
between the rest--frame UV emission  and the 4000 {\AA} break 
is a function of the age of the galaxy.

We have considered two extreme models for the SFR,
namely a single burst with an exponential decline on a time scale
$\tau=10^7$ yr and a constant SFR. The conclusions do not change
within the redshift range $4.4$--$4.7$.
The UV luminosity, as observed in the I--band, $L_{1550}\simeq 
2\times 10^{41}$ erg
s$^{-1}$ {\AA}$^{-1}$, corresponds to a SFR $\approx 30 M_{\odot}$ yr$^{-1}$
in both cases. The rest-frame blue magnitude is
$M_B\sim -21$, of the order of the characteristic magnitude $M^*_B$ of
the local luminosity function of blue galaxies.

From the $I-K\sim 1.3$ colour we deduce an age $\magcir 10^7$ and $\approx 
10^8$ yr for the single burst and the constant SFR case respectively. 
In both cases the converted mass in stars  
is $M_{star} \sim 1-3\times 10^9 M_{\odot}$.

If the galaxy is associated with one of the absorption systems, interesting
estimates of the total gaseous  mass  $M_{gas}$  could be obtained.
For instance, if it is associated with the DLA, where 
almost all of the hydrogen is in  the neutral form, $M_{gas}$  can be estimated
from the observed HI column density $N_{H} = 5\times 10^{20}$ cm$^{-2}$
by assuming a radius in the range 
$13 - 30$ kpc, where the highest value is the upper limit in the impact 
parameter derived from known DLA (Steidel 1995).
The resulting mass is in the range $M_{gas} = 2 - 9 \times 10^9 M_{\odot}$
respectively, implying a total baryonic mass not substantially higher than 
$10^{10} M_{\odot}$ and a gas/stars mass ratio of 2 or larger. 
The efficiency of conversion of cold gas into stars is an 
important free parameter for galaxy formation models:
at the constant rate of 30 $M_{\odot}$ yr$^{-1}$, 
no more than 20\% of the  gaseous mass is converted into stars within
a dynamical time in agreement with
models that reproduce the observed properties
of local galaxies (Kauffman, White \& Guiderdoni 1993).

This galaxy might otherwise be not associated with any
identified absorption systems, and in particular it might be in the
environment of the quasar itself, which might be a preferred site
for galaxy formation. 
Indeed, extended line--emitting companions of
high redshift QSOs have been detected in a number of cases \cite{DS,HS}.
In most cases, they emit  no continuum radiation,
which has been  interpreted as an indication of  an ongoing interaction
with the quasar itself.
Since  all these objects are associated with radio--loud quasars, a
possible connection with the same phenomena
giving rise to the radio emission has been suggested. 
As shown in the previous section, a strong Lyman$\alpha$ emission
from this galaxy might reconcile its colors with those expected
for a $z=4.7$ galaxy: however, it is difficult to ascribe
the  overall 
photometric properties of this galaxy as the result of reprocessing
of the QSO radiation or interaction with it.
Indeed, the clear detection in the I band, 
where no strong emission line is present at $z=4.7$, 
provides the strongest evidence that this objects has a truly UV continuum,
possibly associated with an intense
star--formation activity, which dominates any QSO reprocessed radiation.
The strong Lyman$\alpha$ needed in this case is still compatible with
that expected from a galaxy forming stars at a rate of 
30 $M_{\odot}$ yr$^{-1}$ (Charlot \& Fall, 1993), though the 
large uncertainities 
in the models prevent an accurate comparison.
Furthermore, the absence of radio emission from this QSO \cite{MO}
argues against
the existence of high energy phenomena on several kpc scale, as typical
in   radio--loud objects.
Our galaxy  would therefore be similar to the companion of
the radio--quiet $z=2.75$ quasar Q1548+0917 \cite{SS}. In this case a  
Lyman$\alpha$ emitting  galaxy at a redshift only $1000$ km s$^{-1}$
greater than that of the QSO was detected 5'' apart from it.
Also in this object, indeed, the detection of the UV continuum 
and the absence of high--ionization emission lines from the galaxy
argued against any direct connection with the QSO activity.

Whatsoever the redshift of this galaxy is, we conclude that its
photometric properties are probably due to star--formation activity,
and thus may represent an example of a high redshift galaxy undergoing its
initial burst of star formation.
An upper limit of the order of $10^8$ yr for the age of the galaxy
implies that the bulk of the stellar population formed at $z<6$.
This limit is not far from that of the furthest  QSOs known, and 
could be representative of the epoch of the first galaxy formation.

{\bf Acknowledgments}\\
We thank C.Hazard, M.J.Irwin and R.G. McMahon for
providing the QSO coordinates in
advance of publications, G. Djorgovski for communicating his estimate of
the K magnitude of the galaxy, A. Bressan for providing
the colours of an elliptical galaxy as a function of redshift, and
M. Vietri for  useful discussions.
We are indebted to P. Madau for computing the redshifted spectrum of
the star forming galaxy.
EG acknowledges partial financial support from ASI.

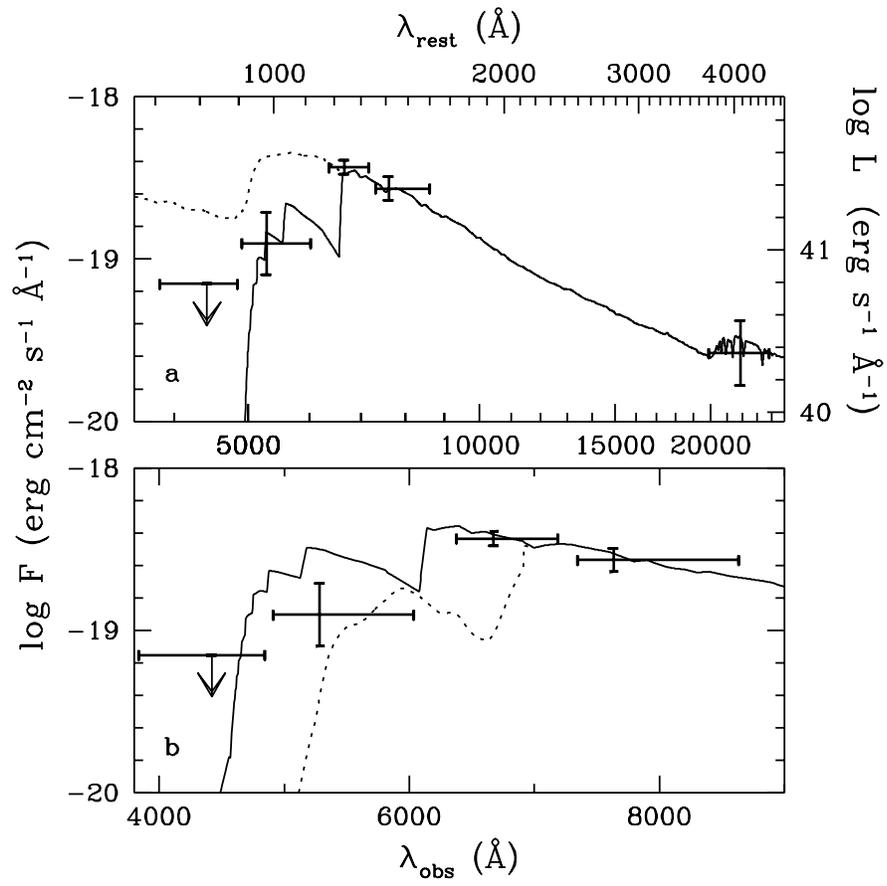
\end{document}